\documentstyle[12pt]{article}

\begin{document}

\font\ftitle=cmbx10 scaled\magstep2
\def\var{\varepsilon}
\def\ov{\overline}
\def\und{\underline}
%baga1030.tex\bigskip\bigskip

\thispagestyle{empty}
 
\vspace*{1cm}

\begin{center}
{\Large \bf The stochastic limit of the Fr\"ohlich Hamiltonian: relations with the quantum Hall effect}
\vspace{7mm}\\
{\large L. Accardi}
\vspace{3mm}\\
Centro Vito Volterra, Universit\`a degli Studi di Roma ``Tor Vergata''\\
via di Tor Vergata, snc, 00133 Roma\\
e--mail: accardi@volterra.mat.uniroma2.it\\
\vspace{3mm}
{\large F. Bagarello}
%\footnote[1]{ Dipartimento di Matematica ed Applicazioni, 
%Fac.Ingegneria, Universit\`a di Palermo, I - 90128  Palermo, Italy}  
\vspace{3mm}\\
  Dipartimento di Matematica ed Applicazioni, 
Fac.Ingegneria, Universit\`a di Palermo, \\I - 90128  Palermo, Italy\\
E-mail: bagarell@unipa.it
\vspace{2mm}\\
\end{center}

\vspace*{5mm}

\begin{abstract}
\noindent 
We propose a model of an approximatively two--dimensional electron gas
 in a uniform electric and magnetic field and interacting with a positive background
through the Fr\"ohlich
Hamiltonian. We consider the stochastic limit of this model and we find the
quantum Langevin equation and the generator of the master equation.
This allows us to calculate the explicit form of the conductivity and
the resistivity tensors and to deduce
 a {\it fine tuning condition} (FTC) between the electric and the
magnetic fields. This condition shows that the $x$--component of the
current is zero unless a certain quotient, involving the physical
parameters, takes values in a finite set of physically meaningful rational numbers. We argue that this behaviour is quite similar to that observed in the quantum Hall effect. We also show that, under some conditions on the form factors entering in the definition of the model, also the plateaux and the "almost" linear behaviour of the Hall resistivity can be recovered. Our FTC does not distinguish between fractional and integer values.
\end{abstract}

\vspace{1cm}

\vfill

\newpage

\section{Introduction }

The Hamiltonian for the quantum Hall effect (QHE) is, see for 
instance reference \cite{BMS},  
\begin{equation}
H^{(N)}=H^{(N)}_0+\lambda(H^{(N)}_c+H^{(N)}_B)\label{6.1}
\end{equation}
where $H^{(N)}_0$ is the Hamiltonian for the free $N$ electrons,
$H^{(N)}_c$ is the  Coulomb interaction:
\begin{equation}
H^{(N)}_c={1\over2}\,\sum^N_{i\not=j}{e^2\over|\und r_i- \und r_j|}\label{6.4}
\end{equation}
and $H^{(N)}_B$ is the interaction of the charges with the positive uniform background.

In the present paper we consider a model defined by an Hamiltonian
\begin{equation}
H=H_{0,e}+H_{0,R}+\lambda H_{eb}
\end{equation}

which is obtained from the Hamiltonian (\ref{6.1}) by introducing the
following approximations (for a more precise description of  $H$ see the next section):

$\bullet$ the Coulomb background-background interaction is 
replaced by the free bosons Hamiltonian $H_{0,R}$, (\ref{37});

$\bullet$ the Coulomb electron-electron and electron-background interaction 
is replaced by the Fr\"ohlich Hamiltonian  $H_{eb}$ (\ref{35})  
which is only quadratic rather than quartic in the fermionic operators.

These are certainly strong approximations. However since, as explained 
in \cite{Str}, from the Fr\"ohlich Hamiltonian it is possible, with a 
canonical transformation, to recover a quartic interaction, 
one can say that the Fr\"ohlich Hamiltonian describes an effective 
electron-electron interaction which may mimic at least some aspects of the 
original Coulomb interaction. From this point of view it seems natural 
to conjecture that some dynamical phenomena deduced from this Hamiltonian 
might have implications in the study of the real QHE. There exists a huge bibliography concerning QHE. Here we refere only to \cite{cha} and, for a more recent review, to \cite{gir}.

This conjecture is supported by our main result, given by formulae 
(\ref{120}) and (\ref{121}) where we deduce,
directly from the dynamics, and not from phenomenological arguments,   
an obstruction to the presence of a non zero $x$-component of the current, 
which is quantized according to the values of a finite set of rational 
numbers.  This result is, to our knowledge, new and the fact that such a 
mechanism can arise even in relatively idealized models of electrical 
conductivity, seems to be at least worth of some attention. More
precisely we prove that the $x$-component of the
mean value of the density current operator is necessarily zero unless
a certain quotient (${2\pi e E\over m\omega^2 L_x}$, cf.
(\ref{12}) and (\ref{15}) for the definition of these parameters),
involving the magnitudes of the physical
quantities defining the model,  takes
a rational value. This is what we call a {\it fine tuning condition} (FTC). 

The rational numbers that appear in the FTC are quotients of the Bohr 
frequencies of the free single--electron Hamiltonian. It is quite reasonable to 
expect that, in a concrete physical situation, only a small number 
of these frequencies will play a relevant role for the scale of 
phenomena involved. In this approximation we can say that the $x$-component 
of the mean value of the density current operator is non zero only in 
correspondence of a finite number of rational values of the fine tuning 
parameter. This feature will be discussed in Section 6.

The fine tuning condition strongly reminds the rational values of the 
filling factor  for which the plateaux 
are observed in the real QHE. Again, in Section 6 we will relate these two facts.

The specification of the values of these physically relevant rational 
numbers and the comparison with those rational numbers which are 
experimentally measured in the QHE, requires a detailed analysis which 
will be done elsewhere.

%Another indication, supporting the above mentioned conjecture, is given 
%by the asymptotic values of the diagonal ($\rho_{xx}$) and off--diagonal
%($\rho_{xy}$) terms, of the resistivity tensor, for large values of the 
%magnetic field (which is the typical experimental situation). 
%For large values of $B=|\vec B|$, $\rho_{xy}$ increases linearly in $B$ 
%while $\rho_{xx}$ is independent of the magnetic field. These results 
%reflect the behaviour experimentally observed in the analysis of the QHE.

We use the technique of the stochastic limit of quantum theory and we 
refer to the paper \cite{AcKoVo96} for a synthetic description, to 
\cite{AcKoVo98b} for more recent results, to \cite{acmat} for mathematical details and to \cite{AcLuVo00} for a  systematic exposition.

\section{The single electron problem}

In these notes we discuss a model of $N<\infty$ charged interacting particles
concentrated around a two dimensional layer contained in the $(x,y)$-plane and
subjected to a uniform electric field $\und E=E\hat j$, along $y$, and to an
uniform magnetic field $\und B=B\hat k$ along $z$.

The Hamiltonian for the free $N$ electrons
$H^{(N)}_0$, is the sum of $N$ contributions:
\begin{equation}
H^{(N)}_0=\sum^N_{i=1}H_0(i)\label{2}
\end{equation}
where $H_0(i)$ describes the minimal coupling of the $i$--th
electrons with the field:
\begin{equation}
H_0(i)={1\over2m}\,\left(\und p+{e\over c}\,\und A(r_i)\right)^2+e
\und E\cdot\und r_i\label{3}
\end{equation}

To $H^{(N)}_0$ we still have to add the interaction with the background and,
then, the free Hamiltonian for the background itself. This will be made in
the following section.

We fix the Landau gauge $\und A=-B(y,0,0)$. In this gauge the Hamiltonian
becomes
\begin{equation}
H_0={1\over2m}\,\left[\left(p_x-{eB\over c}\,y\right)^2+p^2_y
\right]+eEy\label{3a}
\end{equation}
which, obeys the commutation rule $[p_x,H_0]=0$.
The solutions of the eigenvalue equation for the single charge
Hamiltonian (\ref{3a})
\begin{equation}
H_0\psi_{np}(\und r)=\varepsilon_{np}\psi_{np}(\und r),\qquad\qquad
n\in{\bf N},\ p\in{\bf Z}\label{5}
\end{equation}
(where the double index is due to the fact that, two quantum
numbers are necessary to fix the eigenstate) are known, [7], to be of the form:
$\psi(\und r)=C e^{ikx}\varphi(y),$ where $C$ is a normalization
constant fixed by the geometry of the system.
Using this factorization, the time independent Schr\"odinger equation
(\ref{5}) can be rewritten as an harmonic oscillator equation
\begin{equation}
\left({1\over2m}p_y^2+{1\over2}m\omega^2(y-y_0)^2\right)
\varphi(y)=\varepsilon'\varphi(y)\label{9}
\end{equation}
depending on the parameters
\begin{equation}
\omega={eB\over mc  }\ ;\quad \varepsilon'=\varepsilon-{\hbar^2k^2
\over2m}\,+{1\over2}\,m\omega^2y^2_0\ ;\quad
y_0={1\over m\omega^2}\,(\hbar k\omega-eE)\ ;\quad \varepsilon=
{k^2\over2m}\,-{1\over2}\,m\omega^2y^2_0\label{12}
\end{equation}
where $k$ is the momentum along the $x$--axis.
If we  require periodic boundary condition on $x$,
$\psi(-L_x/2,y)=\psi(L_x/2,y)$, for almost all $y$, we also  conclude that
the momentum $k$ along $x$, cannot take arbitrary values but must be
quantized. In particular, if the system is infinitely extended along
$y$, then all the possible values of $k$ are:
\begin{equation}
k={2\pi\over L_x}\,p\ ,\qquad p\in{\bf Z}\label{13}
\end{equation}
Normalizing the wave functions  in the strip
$[-L_x/2,L_x/2]\times{\bf R}$,
we finally get:
\begin{equation}
\psi_{np}(\und r)={e^{i{2\pi px\over L_x}}\over\sqrt{L_x}}\,
\varphi_n(y-y^{(p)}_0)\quad\quad\varepsilon_{np}=\hbar\omega(n+1/2)-{eE
\over2m\omega^2}\,\left(eE-{4\hbar\omega\pi p\over L_x}\right)\label{15}
\end{equation}
where $\varphi_n$ is the $n$--th eigenstate of the one-dimensional
harmonic oscillator, $\omega$ and $y_0$ are given by (\ref{12}) and $p$
is fixed by (\ref{13}).

Equation (\ref{15}) shows that the wave function $\psi_{np}(\und r)$
factorizes in a $x$--dependent part, which is labelled by the
quantum number $p$, and a part,  only depending on $y$, which is labelled
by both $n$ {\bf and\/} $p$ due to the presence of $y^{(p)}_0$ in the argument
of the function $\varphi_n$.

It may be interesting to remark that when $E=0$  the model collapses
to the one of a simple harmonic oscillator, see \cite{gir} and \cite{BMS}
for instance, and
an infinite degeneracy in $p$ of each Landau level ($n$ fixed) appears.
Following the usual terminology we will call {\em lowest Landau level}
(LLL) the energy level corresponding to $n=0$.

\section{The second quantized model}

The Hamiltonian $H^{(N)}_0$ contains the interaction of the electrons with
the electric and the magnetic field. In this paper we add the Fr\"ohlich
interaction of the electrons with a background bosonic field. The free
Hamiltonian of the background boson field is
\begin{equation}
H_{0,R}=\int\omega(\und k)b^+(\und k)b(\und k)d k\label{37}
\end{equation}
where $\omega(\und k)$ is the dispersion  for the free background.
Its analytical form will be kept general in this paper.

The electron--background interaction is given here by the Fr\"ohlich
Hamiltonian \cite{Str}
\begin{equation}
H_{eb}=\int\psi^\dagger(\und r)\psi(\und r)\tilde F(\und r-\und r')\phi(\und
r')d rd r'\label{33}
\end{equation}
where $\psi(\und r)$ and $\phi(\und r')$ are respectively the electron and
the bosonic fields, while $\tilde F$ is a form factor. Expanding $\phi
(\und r)$ in plane waves, $\psi(\und r)$ in terms
of the eigenstates $\psi_\alpha(\und r)$, see (\ref{15}), introducing
the form factors
\begin{equation}
g_{\alpha\beta}(\und k):={1\over\sqrt{(2\pi)^3}}\,{\hat V_{\alpha
\beta}(\und k)\over\sqrt{2\omega(\und k)}}\label{34}
\end{equation}
where
\begin{equation}
\hat V_{\alpha\beta'}(\und k):=\int\overline{\psi_\alpha(\und r)}
e^{i\und k\cdot\und r}\psi_{\beta'}(\und r)d r\label{19}
\end{equation}
and taking
$\tilde F(\und r)=e^2 \delta(\und r)$, \cite{Str}, we can write
\begin{equation}
H_{eb}=e^2 \sum_{\alpha\beta}a^+_\alpha a_\beta(b(g_{\alpha\beta})+b^+
(\ov{g_{\beta\alpha}}))\label{35}
\end{equation}
which is quadratic in the fermionic operators
$a_\alpha$, $a^+_\alpha$ are fermionic operators satisfying
\begin{equation}
\{a_\alpha,a_\beta\}=\{a^+_\alpha,a^+_\beta\}=0\qquad\{a_\alpha,a^+_\beta\}=
\delta_{\alpha\beta}\label{20}
\end{equation}
The boson operators $b(\und k)$ satisfy the canonical comutation
relations:
\begin{equation}
[b(\und k),b^+(\und k')]=\delta(\und k-\und k')\quad[b(\und k),b(\und
k')]=[b^+(\und k),b^+(\und k')]=0
\label{ccr}
\end{equation}
The form factors $g_{\alpha\beta}$ depend on the level indices
$(\alpha,\beta)$. Notice
that we have adopted here and in the following the simplifying notation
for the quantum numbers $\alpha=(n_\alpha,p_\alpha)$ and that we have
introduced the smeared operators
\begin{equation}
b(g_{\beta\alpha})=\int dk\, b(\und k)\,g_{\beta\alpha}(\und k).
\end{equation}

In terms of the fermion operators, the free electron Hamiltonian
(\ref{2}) becomes:
\begin{equation}
H_{0,e}=\sum_\alpha\var_\alpha a^+_\alpha a_\alpha,\label{36}
\end{equation}
where the $\varepsilon_\alpha$ are the single electron energies, labeled
by the pairs $\alpha=(n,p)$ as explained in formula (\ref{15}).

Therefore the total Hamiltonian is:
\begin{equation}
H=H_{0,e}+H_{0,R}+\lambda H_{eb}=H_0+\lambda H_{eb}\label{38}
\end{equation}

\section{The stochastic limit of the model}

In this section we briefly outline how to apply the stochastic limit procedure to
the model introduced above. The stochastic limit describes the
dominating contribution to the dynamics in time scales of the order
$t/\lambda^2$, where $\lambda$ is the coupling constant.
The {\it stochastic golden rule\/} is a prescription which, given a
usual Hamiltonian equation, allows to write, with a few simple
calculations, the Langevin and the master equation, \cite{acmat,AcKoVo96,AcLuVo00}. In this paper we will be mainly concerned with the
master equation. \bigskip

The starting point is the Hamiltonian (\ref{38}) together with the commutation
relations (\ref{20}), (\ref{ccr}).
Of course, the Fermi and the Bose operators commute.
The interaction Hamiltonian $H_{eb}$ for this model is given by
(\ref{35}) and the free Hamiltonian $H_0$ is given by (\ref{36}), (\ref{37}) and (\ref{38}).

The time evolution of $H_{eb}$, in the interaction picture is then
\begin{equation}
H_{eb}(t)=e^{iH_0t}H_{eb}e^{-iH_0t}=e^2\sum_{\alpha\beta}a^+_\alpha
a_\beta(b(g_{\alpha\beta}e^{-it(\omega-\var_{\alpha\beta})})+
b^+(\ov g_{\beta\alpha}e^{it(\omega-\var_{\beta\alpha})}))\label{75}
\end{equation}
where
\begin{equation}\var_{\alpha\beta}=\var_\alpha-\var_\beta\label{43}
\end{equation}

Therefore the Schr\"odinger equation in interaction
representation is:
\begin{equation}
\partial_tU^{(\lambda)}_t=-i\lambda H_{eb}(t)U^{(\lambda)}_t\label{47}
\end{equation}
After the time rescaling $t\to t/\lambda^2$, equation (\ref{47}) becomes
\begin{equation}
\partial_tU^{(\lambda)}_{t/\lambda^2}=-{i\over\lambda} H_{eb}(t/\lambda^2)
U^{(\lambda)}_{t/\lambda^2}\label{48}
\end{equation}
whose integral form is
\begin{equation}
U^{(\lambda)}_{t/\lambda^2}=1\!\!\!\!1-{i\over\lambda}\,\int^t_0H_{eb}
(t'/\lambda^2)U^{(\lambda)}_{t'/\lambda^2}dt'\label{49}
\end{equation}

We see that the rescaled Hamiltonian
\begin{equation}
{1\over\lambda}\,H_{eb}(t/\lambda^2)=e^2\,\sum_{\alpha \beta} a_\alpha^\dagger
a_\beta{1\over
\lambda}\,b\left(e^{-it\over\lambda^2}\,(\omega-\varepsilon_{\alpha \beta})
g_{\alpha \beta}\right)+\hbox{ h.c.}\label{50}
\end{equation}
depends on the rescaled fields
\begin{equation}
b_{\alpha \beta,\lambda}(t)={1\over\lambda}\,b(e^{-i{t\over\lambda^2}\,
(\omega-
\varepsilon_{\alpha \beta})}g_{\alpha \beta})\label{51}
\end{equation}
The first statement of the stochastic golden rule see
\cite{AcLuVo00} is that the rescaled fields
converge (in the sense of correlators) to a quantum white noise
\begin{equation}
b_{\alpha\beta}(t)=\lim_{\lambda\to0}{1\over\lambda}\,b(g_{\alpha \beta}
e^{-i{t\over\lambda^2} \,(\omega-\var_{\alpha \beta})})\label{52}
\end{equation}
characterized by the following commutation relations
\begin{equation}
[b_{\alpha\beta}(t),b_{\alpha'\beta'}(t')]=[b^+_{\alpha\beta}(t),
b^+_{\alpha'\beta'}(t')]=0
\label{82zz}
\end{equation}
\begin{equation}
[b_{\alpha\beta}(t),b^+_{\alpha'\beta'}(t')]=
\delta_{\var_{\alpha\beta},\var_{\alpha'\beta'}}\delta(t-t')
G^{\alpha\beta\alpha'\beta'}
\label{82zzz}
\end{equation}
where the constants $G^{\alpha\beta\alpha'\beta'}$ are given by
\begin{equation}
G^{\alpha\beta\alpha'\beta'}=\int^\infty_{-\infty}d\tau\int d kg_{\alpha
\beta}(\und k)\overline{g_{\alpha'\beta'}(\und
k)}e^{i\tau(\omega(\und k)-\epsilon_{\alpha\beta})}=2\pi\int dk
g_{\alpha\beta}(k)g_{\alpha\beta}(k)\delta(\omega(\und k)-\varepsilon_{\alpha
\beta})\label{78}
\end{equation}
will be denoted by $\eta_0$. The vacuum of the master fields $b_{\alpha\beta}(t)$
\begin{equation}
b_{\alpha\beta}(t)\eta_0=0\quad\forall\,\alpha \, \beta,\ \forall\,t\label{54}
\end{equation}
Moreover the appeareance of $\delta_{\var_{\alpha\beta},\var_{\alpha'
\beta'}}$ in the commutator (\ref{82zzz}) and of the $\delta$--function in
(\ref{78}) is a first indication
of the relevance of the integer numbers  for this model. This point will
be better clarified in the following and will be relevant in the
computation of the conductivity tensor.

The limit Hamiltonian is, then, see \cite{AcLuVo00}.
\begin{equation}
H^{(sl)}_{eb}(t)=e^2\sum_{\alpha\beta}(a^+_\alpha a_\beta b_{\alpha
\beta}(t)+\hbox{ h.c.})\label{76}
\end{equation}
In this sense we say that $H^{(sl)}_{eb}(t)$ is the ``stochastic limit''
of $H_{eb}(t)$ in (\ref{75}).
Moreover, the stochastic limit of the equation of motion is (\ref{48})
\begin{equation}
\partial_tU_t=-iH^{(sl)}_{eb}(t)U_t\label{56}
\end{equation}
or, in integral form,
\begin{equation}
U_t=1\!\!\!\!1-i\int^t_0H^{(sl)}_{eb}(t')U_{t'}dt',\label{57}
\end{equation}

Finally, the stochastic limit of the (Heisenberg) time evolution of
any observable $X$ of the system is:
\begin{equation}
j_t(X)=U^+_t XU_t=U^+_t(X\otimes1_R)U_t\label{58}
\end{equation}
Since the $b_{\alpha\beta}(t)$ are quantum white noises, equation (\ref{56}), and the
corresponding differential equation for $j_t(\tilde X)$, are singular
equations and to give them a meaning we bring them in normal form.
This normally ordered evolution equation is
called {\it the quantum Langevin equation\/}. Its explicit form is:
$$\partial_tj_t( X)=e^2\sum_{\alpha\beta}\{j_t([a^+_\alpha
a_\beta, X]\Gamma^{\alpha\beta}_--\Gamma^{\alpha\beta}_-
[a^+_\beta a_\alpha, X])\}+$$
\begin{equation}
+ie^2\sum_{\alpha\beta}\{b^+_{\alpha\beta}(t)j_t([a^+_\beta
a_\alpha, X])
+j_t([a^+_\alpha a_\beta, X])b_{\alpha\beta}(t)\}\label{80}
\end{equation}
where
\begin{equation}
\Gamma^{\alpha\beta}_-:=\sum_{\alpha'\beta'}\delta_{\varepsilon_{\alpha
\beta},\varepsilon_{\alpha'\beta'}}a^+_{\beta'}a_{\alpha'}
G^{\alpha\beta\alpha'\beta'}_-\label{81}
\end{equation}
\begin{equation}
G_-^{\alpha\beta\alpha'\beta'}=\int^0_{-\infty}d\tau\int d kg_{\alpha\beta}
(\und k)\overline{g_{\alpha'\beta'}(\und
k)}e^{i\tau(\omega(\und k)-\epsilon_{\alpha\beta})}=\label{77}
\end{equation}
$$={1\over2}\,G^{\alpha\beta\alpha'\beta'}-i\hbox{ P.P. }\int
g^{(k)}_{\alpha\beta}\overline{g^{(k)}_{\alpha'\beta'}}{1\over
\omega_k-\varepsilon_{\alpha\beta}}$$

The master equation is obtained by taking the mean value of (\ref{80}) in
the state
$\eta_0^{(\xi)}=\eta\otimes\xi$, $\xi$ being a generic vector of the
system. This gives
\begin{equation}
\langle\partial_t j_t( X)\rangle_{\eta^{(\xi)}_0}=e^2\sum_{\alpha
\beta}\langle j_t([a^+_\alpha a_\beta, X]\Gamma^{\alpha\beta}_-
-\Gamma^{\alpha\beta+}_-[a^+_\beta a_\alpha,
X])\rangle_{\eta^{(\xi)}_0}\label{82}
\end{equation}
and from this we find for the generator
\begin{equation}
L( X)=e\sum_{\alpha\beta\alpha'\beta'}\delta_{\var_{\alpha\beta},
\var_{\alpha'\beta'}}\{[a^+_\alpha a_\beta, X]a^+_{\beta'}
a_{\alpha'}G^{\alpha\beta\alpha'\beta'}_-
-a^+_{\alpha'}a_{\beta'}[a^+_\beta a_\alpha, X]
\overline{G^{\alpha\beta\alpha'\beta'}_-}\}\label{83}
\end{equation}

The expressions for $L( X)$ obtained above will be the starting
point for our successive analysis.

\section{The current operator in second quantization}

The current is proportional to the sum of the
velocities of the electrons:
\begin{equation}
\vec J_\Lambda(t)=\alpha_c\sum^{N}_{i=1}{d\over dt}\,\vec
R_i(t).\label{84}
\end{equation}
Here $\Lambda$ is the two--dimensional region corresponding to the physical
layer, $\alpha_c$ is a proportionality constant which takes into account
the electron charge, the area of the surface of the physical device and
other physical quantities, and $\vec R_i(t)$ is the position operator for
the $i$--th  electron.
Moreover $N$ is the number of electrons contained in $\Lambda$. Defining
\begin{equation}
\vec X_\Lambda(t)=\sum_{i=1}^{N}\vec R_i(t)\ ,\label{85}
\end{equation}
we conclude that
\begin{equation}
\vec J_\Lambda(t)=\alpha_c\dot{\vec X}_\Lambda(t)\ .\label{86}
\end{equation}
Since $\vec X_\Lambda(t)$ is a sum of single-electron operators
its expression in second quantization is given by
\begin{equation}
\vec X_\Lambda=\sum_{\gamma\mu}\vec X_{\gamma\mu}a^+_\gamma
a_\mu\label{87}
\end{equation}
where
\begin{equation}
\vec X_{\gamma\mu}=\langle\psi_\gamma,\vec X_\Lambda\psi_\mu\rangle=\int
\psi_\gamma(\underline r)\underline r\psi_\mu(\underline
r)d r\label{88}
\end{equation}
Recall that the $\psi_\gamma(\underline r)$ are the
single electron wave functions given by (\ref{15}) and $a_\alpha$ and
$a^+_\alpha$ satisfy the anticommutation relations (\ref{20}).

The next step consists in computing the matrix elements (\ref{88}).
This can be done exactly, due to the known expression for
$\psi_\gamma(\underline r)$, even without restricting the analysis to
the LLL. In fact the two components of $\vec X_{\gamma\mu}$ in (\ref{88})
have the form:
\begin{equation}
X^{(1)}_{\gamma\mu}={1\over L_x}\,\int^{L_x/2}_{-L_x/2}x e^{2\pi
i(p_\mu-p_\gamma)x/L_x}dx\cdot
\int^{+\infty}_{-\infty}\overline{\varphi_{n_\gamma}(y-y_0^{(p
\gamma)}})\varphi_{n_\mu}(y-y^{(p_\mu)}_0)dy\label{89a}
\end{equation}
\begin{equation}
X^{(2)}_{\gamma\mu}={1\over L_x}\int^{L_x/2}_{-L_x/2}e^{2\pi
i(p_\mu-p_\gamma)x/L_x}dx\cdot
\int^{+\infty}_{-\infty}y\,\overline{\varphi_{n_\gamma}(y-y^{(p_\gamma)}_0)}
\varphi_{n_\mu}(y-y^{(p_\mu)}_0)dy\label{89b}
\end{equation}
and these integrations  can be easily performed by making use of the following formulas (cf. \cite{Grad}
and \cite{cohen}):
\begin{equation}
\int^{+\infty}_{-\infty}dx e^{-x^2}H_m(x+y)H_n(x+z)=2^n\sqrt\pi
m!z^{n-m}\cdot L^{n-m}_m(-2yz)\label{90a}
\end{equation}
if $m\leq n$, and
\begin{equation}
\int^{+\infty}_{-\infty}\overline{\varphi_n(y)}y\varphi_m(y)dy=
\sqrt{\hbar\over 2mn}[\sqrt{m+1}\delta_{n,m+1}+\sqrt
m\delta_{n,m-1}]\label{90b}
\end{equation}
where $H_m$ and $L^{n-m}_m$ are respectively Hermite and Laguerre
polynomials. With these ingredients we get
\begin{equation}
X^{(1)}_{\gamma\mu}=(1-\delta_{p_\mu p_\gamma})(-1)^{p_\mu-p_\gamma}
{L_xe^{-y^2_{p_\mu
p_\gamma}}\over2\pi_i(p_\mu-p_\gamma)}\,{\cal L}_{\gamma\mu}\label{91a}
\end{equation}
\begin{equation}
X^{(2)}_{\gamma\mu}=\delta_{p_\mu p_\gamma}\{y^{(p_\gamma)}_0
\delta_{n_\gamma
n_\mu}+\sqrt{\hbar\over2m\omega}(\sqrt{n_\mu+1}\delta_{n_\gamma,n_\mu+1}+
\sqrt{n_\mu}\delta_{n_\gamma,n_\gamma-1})\label{91b}
\end{equation}
where
\begin{equation}
{\cal L}_{\gamma\mu}:=\cases{
\sqrt{2^{n_\gamma}n_\mu!\over2^{n_\mu}n_\gamma!}y^{n_\gamma-n_\mu}_{p_\mu
p_\gamma}L^{n_\gamma-n_\mu}_{n_\mu}(2y^2_{p_\mu p_\gamma})\quad \quad\hbox{ if }
n_\mu\leq n_\gamma\cr
\sqrt{2^{n_\mu}n_\gamma!\over 2^{n\gamma}n_\mu!}(-y_{p_\mu
p_\gamma})^{n_\mu-n_\gamma}L^{n_\mu-n_\gamma}_{n_\gamma}
(2y^2_{p_\mu p_\gamma})\quad \quad\hbox{ if }n_\gamma\leq n_\mu\cr}\label{92}
\end{equation}
\begin{equation}
y_{p_\mu p_\gamma}:=\sqrt{m\omega\over
4\hbar}(y^{(p_\mu)}_0-y^{(p_\gamma)}_0)=
{\pi\over L_x}\,\sqrt{\hbar\over m\omega}(p_\mu-p_\gamma)\label{93}
\end{equation}
Notice that, whenever $p_\mu=p_\gamma$, formula (\ref{91a}) must be interpreted
simply as: $X^{(1)}_{\gamma\mu}=0$.

These results are simpler if we restrict to the
LLL. In this case we have $n_\gamma=n_\mu=0$ and therefore, since
$L^a_0(x)=1$, we simply get
\begin{equation}
X^{(1)}_{\gamma\mu}=(1-\delta_{p_\mu p_\gamma})(-1)^{p_\mu-p_\gamma}
L_x {e^{-y^2_{p_\mu p_\gamma}}\over2 \pi i(p_\mu-p_\gamma)}\label{94a}
\end{equation}
\begin{equation}
X^{(2)}_{\gamma\mu}=y^{(p_\gamma)}_0\delta_{p_\mu
p_\gamma}\label{94b}
\end{equation}
To show how these results can be useful in the computation of the
electron current we start noticing that, if $\varrho$ is a state of the
electron system, then
\begin{equation}
\langle\vec J_\Lambda(t)\rangle_\varrho=\alpha_c\langle{d\over dt}\,\vec
X_\Lambda(t)\rangle_\varrho=\alpha_c\langle L(\vec
X_\Lambda(t))\rangle_\varrho=\alpha_c Tr(\varrho{L(\vec
X_\Lambda(t))})\label{95}
\end{equation}

The vector $\langle\vec J_\Lambda(t)\rangle_\varrho$ will be computed in
the next section for a particular class of states $\varrho$, and we will
use this result to get the expressions for the conductivity tensor and for its inverse, the
resistivity matrix.
\bigskip\medskip

\section{The fine tuning condition and the resistivity tensor}

\medskip
In this section we will use formula (\ref{95}) above in order to obtain the
conductivity and the resistivity tensors. To do this we begin
computing the electric current. We first need to find $ L(\vec
X_\Lambda)$, $L$ being the generator given in (\ref{83}). Since $\vec X_\Lambda=
\vec X_\Lambda^\dagger$, we have
$$
L(\vec X_\Lambda)=L_1(\vec X_\Lambda)+h.c.,
$$
where, as we find after a few computations,
\begin{equation}
L_1(\vec X_\Lambda)=e^2\sum_{\alpha\beta\alpha'\beta',\gamma}
\delta_{\epsilon_{\alpha\beta},\epsilon_{\alpha'\beta'}}G_-^{\alpha\beta
\alpha'\beta'}(\vec X_{\beta\gamma}a^+_\alpha a_\gamma a^+_{\beta'}a_{\alpha'}
-\vec X_{\gamma\alpha}a^+_\gamma a_\beta
a^+_{\beta'}a_{\alpha'})\label{96}
\end{equation}
In the present paper we consider a situation of zero temperature  and we
compute the mean value of $L_1(\vec X_\Lambda)$ on a Fock $N$--particle
state $\psi_I$:
\begin{equation}
\psi_I=a^+_{i_1}\dots a^+_{i_{N_I}}\psi_0, \quad \quad i_k\neq i_l,
\forall k\neq l\label{97}
\end{equation}
where $I$ is a set of possible quantum numbers $(I\subset({\bf N}_o,
{\bf Z}))$, $N_I$ is the number of elements in $I$ and $\psi_0$ is
the vacuum vector of the fermionic operators, $a_\alpha \psi_o=0$ for all
$\alpha$. The order of the elements of $I$ is important to fix
uniquely the phase of  $\psi_I$.
Equation (\ref{95}) gives now
\begin{equation}
\langle\psi_I,\vec
J_\Lambda(t)\psi_I\rangle\mid_{t=0}=\alpha_c\langle\psi_I,L(\vec X_\Lambda)
\psi_I\rangle\label{98}
\end{equation}
Introducing now the characteristic function of the set $I$,
\begin{equation}
\chi_I(\alpha)=\cases{
1\hbox{ if }\alpha\in I\cr
0\hbox{ if }\alpha\notin I,\cr}\label{99}
\end{equation}
we get
\begin{equation}
\langle a_\gamma^\dagger a_\alpha\psi_I,a_{\beta'}^\dagger a_{\alpha'}\psi_I
\rangle=\delta_{\alpha\gamma}\delta_{\alpha'\beta'}\chi_I(\alpha) \chi_I
(\alpha')+\delta_{\alpha\alpha'}\delta_{\gamma\beta'}\chi_I(\alpha)(1-\chi_I
(\gamma)).
\label{100}
\end{equation}

Using this equality, together with
\begin{equation}
\delta_{\var_{\alpha\beta},\var_{\alpha'\alpha'}}=\delta_{\var_{\alpha},
\var_{\beta}}\qquad\qquad\delta_{\var_{\alpha\beta},\var_{\alpha\beta'}}=
\delta_{\var_\beta,\var_{\beta'}}\label{101}
\end{equation}
we find that the average current is proportional to
\begin{equation}
\langle L(\vec X_\Lambda)\rangle_{\psi_I}={\cal L}_1(\vec
X_\Lambda)+{\cal L}_2(\vec X_\Lambda)\label{102}\end{equation}
where we isolate two contributions of different structure:
\begin{equation}
{\cal L}_1(\vec X_\Lambda)=e^2\sum_{\alpha\beta\alpha'}\delta_{\var_\alpha,
\var_\beta}
\{\chi_I(\alpha)-\chi_I(\beta)\}\chi_I(\alpha')
(\vec X_{\alpha\beta}\overline{G_-^{\alpha\beta\alpha'\alpha'}}+\vec
x_{\beta\alpha}G_-^{\alpha\beta\alpha'\alpha'}),
\label{103a}
\end{equation}
$${\cal L}_2(\vec X_\Lambda)=e^2\sum_{\alpha\beta\beta'}\delta_{\var_\beta,
\var_{\beta'}}\{\vec X_{\beta\beta'}[G_-^{\alpha\beta\alpha\beta'}
\chi_I(\alpha)(1-\chi_I(\beta')) -\overline{G_-^{\beta\alpha\beta'\alpha}}\chi_I(\beta')(1-\chi_I(\alpha))]-$$
\begin{equation}
-\vec X_{\beta'\beta}[G_-^{\beta\alpha\beta'\alpha}\chi_I(\beta')
(1-\chi_I(\alpha)) -\overline{G_-^{\alpha\beta\alpha\beta'}}\chi_I(\alpha)
(1-\chi_I(\beta'))]\}.\label{103b}
\end{equation}
\bigskip

\noindent{\sc Remark}.
It is interesting to notice that if we replace $\delta_{\var_\alpha,
\var_\beta}$ by $\delta_{\alpha,\beta}$ and $\delta_{\var_\beta,
\var_{\beta'}}$ by $\delta_{\beta,\beta'}$, then we easily obtain $\langle
L( X_\Lambda^{(1)})\rangle_{\psi_I}=0$, which would imply that no current
transportation is compatible with this constraint. This means that this
approximation (taking $\alpha=\beta$ and $\beta=\beta'$
means to consider only one among the many contributions in the sums in
(\ref{103a}), (\ref{103b})!) is too strong and must be avoided in order not
to get trivial results.\bigskip

Using equations (\ref{91a}), (\ref{91b}) for $X^{(i)}_{\gamma\mu}$ we are able
to obtain ${\cal L}_1(X^{(i)}_\Lambda)$ and ${\cal L}_2(X^{(i)}_\Lambda)$,
$i=1,2$. First of all we can show that, even if ${\cal L}_1(X^{(1)}_\Lambda)$
is not zero, nevertheless it does not depend on the electric field.
Therefore
\begin{equation}
{\partial\over \partial E}{\cal L}_1(X^{(1)}_\Lambda)=0\label{104}
\end{equation}
Secondly, the computation of ${\cal L}_2(X^{(1)}_\Lambda)$ gives rise to
an interesting phenomenon: due to the definition of $X^{(1)}_{\gamma\mu}$,
the sum in (\ref{103b}) is different from zero only if $p_\beta\neq
p_{\beta'}$. Moreover, we also must have
$\var_\beta=\var_{\beta'}$, that is
\begin{equation}
n_\beta-n_{\beta'}={2\pi eE\over
m\omega^2L_x}\,(p_{\beta'}-p_\beta)\label{105}
\end{equation}
This equality can be satisfied in two different ways: let us denote
${\cal R}$ the set of all possible quotients of the form
$(n_\beta-n_{\beta'})/(p_{\beta'}-p_\beta)$. This set, in principle, coincides with the set of the rational numbers. Therefore $0\in{\cal R}$. Then
\begin{itemize}
\item[1)] if ${2\pi eE\over m\omega^2L_x}$ is not in ${\cal R}$,
(\ref{105}) can be satisfied only if $\beta=\beta'$. But this condition implies in
particular that $p_\beta= p_{\beta'}$, and we know already that whenever this
condition holds, then $X_{\beta\beta'}^{(1)}=0$, so that ${\cal L}_2
(X^{(1)}_\Lambda)=0$.
\item[2)] If ${2\pi eE\over m\omega^2L_x}$ is in ${\cal R}$, then we have
two possibilities: the first one is again
$$\beta=\beta'$$
which, as we have just shown, does not contribute to ${\cal L}_2
(X^{(1)}_\Lambda)$. The second is
\begin{equation}
{n_\beta-n_{\beta'}\over p_{\beta'}-p_\beta}\,={2\pi eE\over
m\omega^2L_x}\label{107}
\end{equation}
which gives a non trivial contribution to the current.

Therefore, we can state the following
\end{itemize}\bigskip

\noindent{\sc Proposition}. {\sl In the context of Model (\ref{38})
there exists a set of rational numbers ${\cal R}$ with the following
property: if the electric and the magnetic fields are such that if the
quotient
$${2\pi eE\over m\omega^2L_x}$$
does not belong to ${\cal R}$ then
$$\langle J^{(1)}_\Lambda(t)\rangle_{\psi_I}=0.$$
}
\bigskip

On the other hand, if condition (\ref{107}) is satisfied, we can conclude that
the sum $\sum_{\alpha\beta\beta'}\delta_{\varepsilon_\beta,\varepsilon_{\beta'}}
(\dots)$ in (\ref{103b}) can be replaced by
\begin{equation}
\sum_{\alpha\beta\beta'}\delta_{\varepsilon_\beta,\varepsilon_{\beta'}}
(\dots)=\sum_{\alpha}{\sum_{\beta\beta'}}'(\dots)\label{108}
\end{equation}
where $\sum_\alpha\sum'_{\beta\beta}$ means that the sum is extended to all
the $\alpha$ and to those $\beta$ and $\beta'$ with $p_\beta\neq p_{\beta'}$
satisfying (\ref{107}) (which automatically implies that $\varepsilon_\beta=
\varepsilon_{\beta'}$).

Since, as it is easily seen,
$g_{\alpha\beta}(k)\overline{g_{\alpha'\beta'}(\underline k)}$
does not depend on $\vec E$, we find that
\begin{equation}
{\partial\over \partial E}G^{\alpha\beta\alpha'\beta'}_-=-i{he\over m\omega
L_x}\,(p_\alpha-p_\beta)\Lambda^{\alpha\beta\alpha'\beta'}_-\label{109}
\end{equation}
where
\begin{equation}
\Lambda^{\alpha\beta\alpha'\beta'}_-=\int^0_{-\infty}d\tau
\int d kg_{\alpha\beta}(\underline
k)\overline{g_{\alpha'\beta'}(\underline k)}e^{i\tau(\omega(\und
k)-\var_{\alpha\beta})}\label{110}
\end{equation}
so that, using also (\ref{108}), we get
\begin{equation}
{\partial\over \partial E}{\cal L}_2(X_\Lambda^{(1)})={he\over m\omega
L_x}\Theta_x \label{(111)}
\end{equation}
where
$$\Theta_x:=\sum_{\alpha}{\sum_{\beta\beta'}}'(p_\beta-p_\alpha)
\tilde x^{(1)}_{\beta\beta'}\{\chi_I(\alpha)(1-\chi_I(\beta'))\cdot
(\Lambda^{\alpha\beta\alpha\beta'}_-+\ov{\Lambda^{\alpha\beta
\alpha\beta'}_-})$$
\begin{equation}
-\chi_I(\beta')(1-\chi_I(\alpha))
(\Lambda^{\beta\alpha\beta'\alpha}_-+\ov{\Lambda^{\beta\alpha\beta'
\alpha}_-})\}\label{112}
\end{equation}
and
\begin{equation}
\tilde x^{(1)}_{\beta\beta'}=i \, X^{(1)}_{\beta\beta'}\quad(\in{\bf R})\label{113}
\end{equation}
Therefore we conclude that
\begin{equation}
{\partial\over \partial E}\langle
J^{(1)}_\Lambda(t)\rangle_{\psi_I}={\alpha_c he^3\over m\omega L_x}\,
\Theta_x\label{114}
\end{equation}

Let us now compute the second component of the average current: $\langle\psi_I,
L(X^{(2)}_\Lambda)\psi_0\rangle={\cal L}_1(X^{(2)}_\Lambda)+{\cal L}_2
(X^{(2)}_\Lambda)$.

The first contribution is easily shown, from (\ref{103a}) and
(\ref{91b}), to be identically zero, since
\begin{equation}
\delta_{\varepsilon_\alpha,\varepsilon_\beta}\delta_{p_\alpha p_\beta}
=\delta_{\alpha\beta}\label{115}
\end{equation}

On the contrary the second term, ${\cal L}_2(X^{(2)}_\Lambda)$,
is different from zero and it has
an interesting expression: in fact, due to the factor
$\delta_{p_\mu, p_\gamma}$, the only non trivial
contributions in the sum $\sum_{\beta\beta'}\delta_{\var_\beta,
\var_{\beta'}}$, in (\ref{103b}), are exactly those with $\beta=\beta'$.
Taking all this into account, we find that
\begin{equation}
{\cal L}_2(X^{(2)}_\Lambda)=e^2\sum_{\alpha\beta}(y^{(p_\beta)}_0-y^{(p_\alpha)}_0)
\chi_I(\alpha)(1-\chi_I(\beta))
(G^{\alpha\beta\alpha\beta}_-+\ov{G^{\alpha\beta\alpha\beta}_-})\label{116}
\end{equation}
which is different from zero. Furthermore, using (\ref{109}), we get
$${\partial\over \partial E}{\cal L}_2(X^{(2)}_\Lambda)=-2e^3\left({h\over m
\omega L_x}\right)^2 \Theta_y$$
were we have defined
\begin{equation}
\Theta_y
=\sum_{\alpha,\beta}(p_\alpha-p_\beta)^2\chi_I(\alpha)(1-\chi_I(\beta))\hbox{
Im }(\Lambda^{\alpha\beta\alpha\beta}_-)\label{117}
\end{equation}
and $\Lambda^{\alpha\beta\alpha\beta}_-$ is given by (\ref{110}). If we call now
$$j_{x,E}={\partial \langle J^{(1)}_\Lambda(t)\rangle_{\psi_I}\over
\partial E} |_{t=0}=\alpha_c {\partial \langle L(X_\Lambda^{(1)})\rangle_{\psi_I}\over
\partial E}$$
$$j_{y,E}={\partial \langle J^{(2)}_\Lambda(t)\rangle_{\psi_I}\over
\partial E} |_{t=0}=\alpha_c {\partial \langle L(X_\Lambda^{(2)})\rangle_{\psi_I}\over
\partial E}\ ,$$
we obtain the conductivity tensor (see \cite{cha})
\begin{equation}
\sigma_{xx}=\sigma_{yy}=j_{y,E}, \quad\quad \sigma_{xy}=-\sigma_{yx}=
j_{x,E}\label{118}
\end{equation}
and the resistivity tensor
\begin{equation}
\rho_{xx}=\rho_{yy}={\sigma_{yy}\over\sigma_{yy}^2+\sigma_{xy}^2},
\quad\quad \rho_{xy}=-\rho_{yx}={\sigma_{xy}\over\sigma_{yy}^2+
\sigma_{xy}^2}\label{119}
\end{equation}
After minor computations we conclude that
\begin{equation}
\rho_{xy}=\cases{
0\qquad\qquad\qquad\qquad\qquad\qquad\hbox{if }{2\pi e E\over m\omega^2 L_x}
\notin{\cal R}\cr
{m\omega L_x\over 2 e^3h \alpha_c}{\Theta_x\over [\Theta_x^2+({h\over m\omega
L_x})^2\Theta_y^2]}\qquad\qquad\hbox{if }{2\pi e E\over m\omega^2
L_x}\in{\cal R},
\cr}\label{120}
\end{equation}
\begin{equation}
\rho_{xx}=\cases{
-({ m\omega L_x\over h})^2{1\over 2 \alpha_c e^3\Theta_y}\qquad\qquad\qquad\qquad\hbox{if }
{2\pi e E\over m\omega^2 L_x}\notin{\cal R}\cr
-{1\over 2 e^3\alpha_c}{\Theta_y\over [\Theta_x^2+({h\over m\omega L_x})^2
\Theta_y^2]}\qquad\qquad\qquad\hbox{if }{2\pi e E\over m\omega^2
L_x}\in{\cal R}\ ,\cr}\label{121}
\end{equation}

We want to relate these results with the experimental graphs concerning the components of the resistivity tensor, see \cite{gir}. To avoid confusions, let us remark that our choice for the direction of the electric field, the $y$ axis, is not the usual one, the $x$ axis, see \cite{gir}. Therefore, in our notation, the Hall resistivity is really $\rho_{xx}$, while our $\rho_{xy}$ corresponds to the $xx$ component of $\rho$ as given in \cite{gir}.

Let us now comment these results which are consequences of the basic
relation (\ref{107}). As it is evident from the formula above, the
fact that the {\it fine tuning condition\/} (FTC) (${2\pi e E\over m\omega^2
L_x}\in{\cal R}$) is satisfied implies that $\rho_{xy}\neq 0$, so that the
resistivity tensor is non-diagonal. Vice-versa, if the FTC is not satisfied,
then $\rho=\rho_{xx}1\!\!1$, $1\!\!1$ being the $2\times 2$ identity matrix.
This implies that, whenever the FTC holds, then the $x$-component of the mean
value of the density current operator is in general different from zero, while
it is necessarely zero if the FTC is not satisfied.

If the
physical system is prepared in such a way that ${2\pi e E\over m\omega^2 L_x}
\in{\cal R}$, then an experimental device should be able to measure a non zero current
along the $x$-axis. Otherwise, this current should be zero whenever ${2\pi e
E\over m\omega^2 L_x}\notin{\cal R}$. A crucial point is now the
determination of the set ${\cal R}$, of rational numbers.
From a mathematical
point of view, all the natural integers $n_\alpha$ and all the relative integer
$p_\alpha$ are allowed. However physics restricts the experimentally
relevant values to a rather small set. In
fact eigenstates corresponding to high values of $n_\alpha$ and
$p_\alpha$ are energetically not favoured because the associated
eigenenergies
$\varepsilon_{n_\alpha p_\alpha}$, in (\ref{15}) increases and the
probabilities of finding an electron in the corresponding eigenstate decrease
(this is a generalization of the standard argument which restrict the analysis
of the fractional QHE to the first few Landau levels).
Moreover, high positive values of $-p_\alpha$ are not compatible with the
fact that $H_0$ must be bounded from below, to be a 'honest' Hamiltonian.

Therefore, in formula (\ref{107}) not all the rational numbers are physically
allowed but only those compatible with the above constraints. For this
reason it is quite reasonable to expect that the set ${\cal R}$ consists
only of a {\it finite set\/} of rational values. The determination of
this set strongly depends on the physics of the experimental setting and
we shall discuss it in a future paper.

We end this section, and the paper, with the following two remarks:

the sharp values of the magnetic field involved in the FTC may be a consequence of the approximation intrinsic in the stochastic limit procedure, which consists in taking $\lambda\rightarrow 0$ and $t\rightarrow \infty$. In intermediate regions ($\lambda> 0$ and $t<\infty$), it is not hard to imagine that the $\delta$-function giving rise to the FTC becames a smoother function.

Under special assumptions on the $B$-dependence of $\Theta_x$ and $\Theta_y$, together with some reasonable physical constraint on the value of the magnetic field, it is not difficult to check that $\rho_{xx}$ has plateaux corresponding to the zeros of $\rho_{xy}$ and that, outside of these plateaux, it grows linearly with $B$.

\vspace{5mm}

\noindent{\bf Acknowledge}. Fabio Bagarello  is grateful to
the Centro Vito Volterra and to CNR for its financial support. The authors
thank Prof. Toyoda for his interesting comments and suggestions.


\vspace{1cm}

\begin{thebibliography}{99}
\bibitem{BMS} {\sc F. Bagarello, G. Marchio, F. Strocchi}, {\it Phys.
Rev.\/} B {\bf 48}, 5306 (1993).
\bibitem{Str} {\sc F. Strocchi}, {\it Elements of quantum mechanics of
 infinite systems\/}, World Scientific, Singapore-Philadelphia.
\bibitem{cha} {\sc T. Chakraborty {\rm and} P. Pietil\"ainen}, {\it The
FQHE\/}, Springer--Verlag, Berlin, 1988.
\bibitem{gir} {\sc S.M. Girvin}, {\it The Quantum Hall Effect: Novel 
Excitations and Broken Symmetries\/}, 
Springer Verlag (1999).
\bibitem{AcKoVo96}
Accardi L. , S.V. Kozyrev, I.V. Volovich
Dynamics of dissipative two--state systems in the stochastic approximation.
Phys. Rev. A 56 N. 3 (1996)
\bibitem{AcKoVo98b}
L. Accardi,  S.V. Kozyrev and I.V.  Volovich:
Dynamical origins of $q$-deformations in QED and the stochastic limit
Journal of Physics A, Math. Gen. 32 (1999) 3485--3495
q-alg/9807137
\bibitem{acmat} Accardi L., Frigerio A., Lu Y.G., Comm. Math. Phys. {\bf 131} (1990) 537-570, Accardi L.,  Lu Y.G., Comm. Math. Phys. {\bf 180} (1960) 605-632
\bibitem{AcLuVo00} {\sc Accardi L.,  Y.G. Lu, I. Volovich},
Quantum Theory and its Stochastic Limit. Springer (2001).
\bibitem{Grad} {\sc I.S. Gradshteyn {\rm and} I.M. Ryzhik}, {\it
Table of Integrals, Series and Products\/}, Academic Press, New York and
London 1980.
\bibitem{cohen} {\sc C.Cohen-Tannoudji, B. Diu, F. Lal$\ddot o$e},
{\em Quantum Mechanics}, John Wiley and Sons, New York (1977).


\end{thebibliography}
\end{document}